\documentclass[review]{elsarticle}

\usepackage{lineno,hyperref}
\modulolinenumbers[5]

\usepackage{amsmath,graphicx,amssymb,amsthm,bbm}
\usepackage{adjustbox,bm,diagbox}
\usepackage{hyperref}

\usepackage{tikz}
\usetikzlibrary{decorations.markings}
\usepackage[caption=false,font=footnotesize]{subfig}
\usepackage{upgreek}
\usepackage{color}
\usepackage{soul}
\usepackage[boxruled]{algorithm2e}
\usepackage[english]{babel}

\journal{Journal of \LaTeX\ Templates}

\begin{document}

\begin{frontmatter}

\title{EEG-assisted Modulation of Sound Sources in the Auditory Scene}


\author[mymainaddress]{Marzieh Haghighi\corref{mycorrespondingauthor}}
\cortext[mycorrespondingauthor]{Corresponding author}
\ead{haghighi@ece.neu.edu}

\author[mymainaddress]{Mohammad~Moghadamfalahi}
\author[mysecondaryaddress]{Murat~Akcakaya}
\author[mymainaddress]{Deniz~Erdogmus}

\address[mymainaddress]{Northeastern University, 360 Huntington Ave, Boston, MA 02115}
\address[mysecondaryaddress]{University of Pittsburgh, 4200 Fifth Ave, Pittsburgh, PA 15260}

\begin{abstract}

Noninvasive EEG (electroencephalography) based auditory attention detection could be useful for improved hearing aids in the future. This work is a novel attempt to investigate the feasibility of online modulation of sound sources by probabilistic detection of auditory attention, using a noninvasive EEG-based brain computer interface. Proposed online system modulates the upcoming sound sources through gain adaptation which employs probabilistic decisions (soft decisions) from a classifier trained on offline calibration data. In this work, calibration EEG data were collected in sessions where the participants listened to two sound sources (one attended and one unattended). Cross-correlation coefficients between the EEG measurements and the attended and unattended sound source envelope (estimates) are used to show differences in sharpness and delays of neural responses for attended versus unattended sound source. Salient features to distinguish attended sources from the unattended ones in the correlation patterns have been identified, and later they have been used to train an auditory attention classifier.  Compared to the existing results in the literature, in this paper we have two main contributions. First, using the auditory attention classifier, we have shown high offline detection performance with single channel EEG measurements of shorter duration compared to the existing approaches in the literature which employ large number of channels with longer EEG measurements. Second, using the classifier trained offline in the calibration session,  we have shown the performance of the online sound source modulation system. We observe that online sound source modulation system is able to keep the level of attended sound source higher than the unattended source.  

\end{abstract}


\end{frontmatter}


\section{Introduction}


Approximately 35 million Americans ($11.3\%$ of the population) suffer from hearing loss; this number is increasing and is projected to reach 40 million by 2025 ~\cite{kochkin2009marketrak}. Within this population only $30\%$ prefer using current generations of hearing aids that are available on the market. One of the most common complaints associated with hearing-aid use is understating speech in the presence of noise and interferences. Effects of interfering sounds on masking the speech intelligibility and audibility have been widely studied~\cite{gelfand2016hearing},~\cite{chung2004challenges}. Specifically, it has been shown that increase in SNR needed for the same level of speech understanding given a background noise for people with hearing loss can be as high as 30 dB more compared to people with normal hearing~\cite{chung2004challenges}. Therefore, amplifying the target signal versus unwanted noises and interferences to facilitate hearing and increase speech intelligibility and listening comfort is one of the basic concepts exploited by hearing aids~\cite{chung2004challenges}. Identifying the signal versus noise is a main step required for the design of a hearing aid. This identification step can be a difficult task in complicated auditory scenes like a cocktail party scenario in which both signal and interferences have acoustic features of speech and can instantly switch their roles based on the attention of the listener and can not be detected based on the predefined assumptions on signal and noise features. Our brain distinguishes the audio sources based on their spectral profile, harmonicity, spectral or spatial separation, temporal onsets and offsets, temporal modulation, and temporal separation~\cite{yost1997cocktail},\cite{yost1993auditory} and focus on one sound to analyse the auditory scene~\cite{bregman1994auditory} in the so called cocktail party effect~\cite{cherry1953some}. Existence of each cue can reduce informational and energetic masking of competing sources and help focusing our attention on the target source. A general overview of the computational efforts in bottom-up or top-down modelling of auditory attention in a cocktail party setting is provided in \cite{kaya2017modelling}.







Brain/Body Computer Interface (BBCI) systems can be used to augment the current generations of hearing aids by discriminating among attended and unattended sound sources. They can be incorporated to provide external evidence based on top-down selective attention of listeners~\cite{ungstrup2014hearing}. Attempts have been made to incorporate bottom-up attention evidences in the design of the hearing aids. Direction based hearing aids that detect attention direction from eye gaze and amplify sounds coming from that direction can be examples of bottom-up attention evidence incorporation~\cite{kidd2013design}. Moreover, there are attempts to use electroencephalography (EEG)-based brain computer interfaces (BCIs) for the identification of attended sound sources. EEG has been extensively used in BCI designs due to its high temporal resolution, non-invasiveness, and portability. These characteristics, in addition to EEG devices being inexpensive and accessible, make EEG a practical choice for the design of a BCI that can be integrated into hearing aids to identify auditory attention.  A crucial step in such an integration is to build an EEG-based BCI that employs auditory attention. 

EEG-based auditory BCIs that rely on external auditory stimulation have recently attracted attention from the research community.  For example, auditory-evoked P300 BCI spelling system for locked-in patients is widely studied ~\cite{schreuder2011listen}, \cite{hohne2011novel}, \cite{kubler2009brain}, \cite{halder2010auditory}, \cite{furdea2009auditory}, \cite{kanoh2008brain}. It was shown that fundamental frequency, amplitude, pitch and direction of audio stimuli are distinctive features that can be processed and distinguished by the brain. Also, more recent studies using EEG measurements have shown that there is a cortical entrainment to the temporal envelope of the attended speech~\cite{ding2014cortical},~\cite{aiken2008human},~\cite{kong2014differential}. A study on the quality of  cortical entrainment to auditory stimulus envelope by top-down cognitive attention has shown enhancement of obligatory auditory processing activity in top-down attention responses when competing auditory stimuli differ in spatial direction ~\cite{power2011endogenous} and frequency~\cite{sheedy2014endogenous}. 

Recently, EEG-based BCI has also been used in cocktail party problems for the classification of attended versus unattended sound sources~\cite{o2014attentional},~\cite{mirkovic2015decoding},~\cite{horton2014envelope}. In the identification of the attended sound source in a cocktail party problem, stimulus reconstruction to estimate the envelope of the input speech stream from high density EEG measurements is the state-of-the-art practice~\cite{pasley2012reconstructing},~\cite{o2014attentional}. In the aforementioned model, envelope of the attended stimulus is reconstructed using spatio-temporal linear decoder applied on neural recordings. In one study that considered the identification of the attended sound source in a dichotic (different sounds playing in each ear) two speaker scenario, 60 seconds of high density EEG data recorded through 128 electrodes were used in the stimulus reconstruction. Two decoders using the attended and unattended speech were trained and it was shown that estimated sound source using the attended decoder has higher correlation with the attended speech envelope compared to the estimated stimuli using unattended decoder with unattended speech envelope~\cite{o2014attentional}. For practical purposes, further studies have attempted to examine the stimulus reconstruction approach using smaller number of EEG electrodes \cite{mirkovic2015decoding} or even two bilaterally placed around the ear electrode arrays (cEEGrids) \cite{mirkovic2016target}. 
Furthermore, the robustness of the attended speech envelope reconstruction in noisy real world acoustic scenes has been demonstrated ~\cite{fuglsang2017noise}. In contrast to the stimulus reconstruction methods, studies with system identification approaches to solve this problem, have tried to reconstruct the neural measurements using the linear forward map of sound sources \cite{machens2004linearity}, \cite{ding2012neural}, \cite{fiedler2017single}, \cite{alickovic2016system}. In a recent related study, a single in-Ear-EEG electrode and an adjacent scalp-EEG electrode were used for auditory attention detection in a diotic two speaker scenario~\cite{fiedler2017single}. On the other hand, in a different class of target speaker detection approaches, studies have tried to extract informative and distinguishable features of EEG measurements with respect to the attended and unattended sound sources to train a classifier~\cite{horton2014envelope},~\cite{haghighi2016audio}. 
In a related study, authors have compared three types of features extracted from speech signal and EEG measurements to learn a linear classifier for the identification of the attended speaker using 20 seconds of data from high density 128 channels EEG recordings~\cite{horton2014envelope}. In our previous related work, we have investigated the role of spectral and spatial features of competing sound sources in an auditory BCI system with the purpose of detecting the attended auditory source in a cocktail party setting. We reported high single channel classification performance for attended sound source versus unattended one based on their spectral and spatial separation of the sources (diotic and dichotic paradigms) using 60 seconds of EEG and stimuli data~\cite{haghighi2016audio}.

In continuation of the above described literature and our previous work, this paper presents two contributions to the literature of EEG-based auditory attention detection in a cocktail party setting:
\begin{itemize}
    \item First, we show successful identification of attended speaker source in a diotic (both sounds playing in both ears) two speaker scenario using 20 seconds of EEG data recorded from 16 channels. The presented classifier outperforms EEG-based auditory attention detectors previously presented in the literature in terms of accuracy, with smaller number of EEG channels (sparse 16 versus typically dense 96 or more), and using time-series of shorter durations (20 seconds versus typically 60 seconds). In fact, using 20 seconds of EEG data from only one of the 16 EEG channels, we demonstrate high classification performance for the auditory attention detection. This extends our results from~\cite{haghighi2016audio}, which showed high single-channel classification accuracy when the EEG duration was 60 seconds. 
    \item Second, we introduce a novel online system that gives feedback on attention of the user in the form of attended to unattended source energy ratio amplification. The level of amplification of attended versus suppression of unattended source is assigned based on a probabilistic model defined over the classifier trained on the offline data including temporal dependency of the user's attention. The goal of the online system is using the probabilistic information of the user's attention to enhance the concentration of the user on the target source in multi-speaker scenarios. The introduced framework for online system is a proof of concept for design perspective of an EEG-augmented hearing aid system. Finally, we show the introduced online system in average is able to keep the level of attended source higher despite statistical changes happening in online data compared to the offline data used for training the classifier.
\end{itemize}

\section{System Overview}
\begin{figure}[h]
        \begin{minipage}[b]{1.0\linewidth}
        \centering
        \includegraphics[width=\columnwidth]{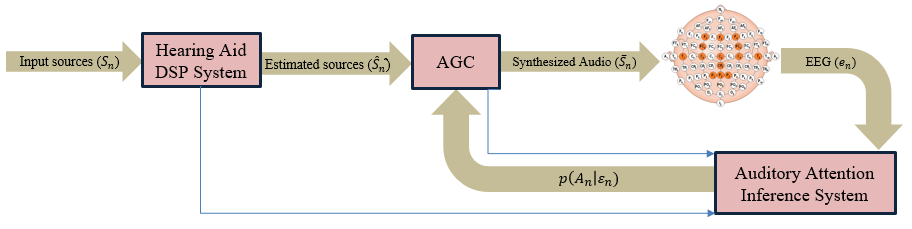}
        \vspace{-15pt}
        \end{minipage}
        \caption{EEG-augmented BCI sytem overview.}
        \label{fig:systemOverview}
    \end{figure}

The diagram represented in Figure~\ref{fig:systemOverview} summarizes the steps of the proposed BCI system. The proposed system gets the mixture of sounds from the environment as the input and modifies the gain of each specific sound. The output of this system is the input to the ear channel. 

The decision on gain modification of each sound is made by the BCI module which consists of three submodules of gain controller, auditory attention inference system and hearing aid DSP system. Hearing aid DSP system estimates independent sound sources from the mixture of sounds in the environment and outputs the information to the gain controller and attention inference module. In this work, we assume that we have the estimated sources which are the outputs of the DSP system based on blind source separation.  

Auditory attention inference system estimates the probability of attention on each specific sound source using EEG measurements and estimated sound sources. Gain controller system takes the estimated probabilities from the attention inference system to modify gains of each specific sound. The details of the attention inference system and gain adjustments are provided in the following sections.\\
    
\subsection{\bf{Online Gain Controller System}}
\label{cha:GC}
Let’s assume that  $S_{n}=(\mathbf{s}_{1,n},...,\mathbf{s}_{i,n},...,\mathbf{s}_{M,n})$  is a matrix containing original sources that each $\mathbf{s}_{i,n}$ is a column vector for $i^{th}$ sound channel for $n^{th}$  round of sending feedback. $\hat{S}_{n}=(\hat{\mathbf{s}}_{1,n},...,\hat{\mathbf{s}}_{i,n},...,\hat{\mathbf{s}}_{M,n})$, would be the estimated source matrix after blind source separation, which we assume exists and its design is out of the scope of this paper. $\mathbf{w_n}=(w_{1,n},...,w_{i,n},...,w_{M,n})^\intercal$ is the vector of weights with $w_{i,n}$ being a scalar showing the gain of $i^{th}$  estimated sound source; and $e_n$ is the EEG evidence vector for $n^{th}$  round. $A_n=i$, indicates the attention of subject is on the $i^{th}$ sound source. Subject will start listening to all sounds with equal energy and then based on brain interface decisions for subject attention on each sound source, speech enhancement or automatic gain controller (AGC) module will assign appropriate weights to each sound source for ${n+1}^{th}$ round of sending feedback according to the following equation:
\begin{equation} \label{eq1}
\mathbf{w_{n+1}}=f(p(A_n|\varepsilon_n(\hat{S}_n,e_n)),\mathbf{w_{n-j}})      
\qquad j=1,2,.. 	
\end{equation}        		         
Equation~\ref{eq1} states that the weights for the upcoming sound sources (${n+1}^{th}$ round) will be decided based on probability of attention given current EEG evidence (${n}^{th}$ round) and previous weights that were used at the ${n-1}^{th}$ round. 
The selection of optimal gain control policies (choosing the form of $f$) that considers other factors such as sound quality due to amplitude modulation, response time to changes versus robustness to outlier incidents influencing brain interface decisions, is anticipated to be a significant and important research area in itself, and we will explore alternative designs in future work.

\subsection{\bf{Auditory Attention Inference System}}
\label{cha:AAIS}
This module calculates probability of attention given EEG evidence. It takes raw EEG measurements, (estimated) sound sources and weights to extract EEG features (evidence), as explained in Section \ref{cha:Methods}. Then, using Bayes rule, the posterior probability distribution of attention over sources is expressed as the product of EEG evidence likelihood times the prior probability distribution over sources,
    \begin{equation}
        \begin{split}
            P(A_n=i|\varepsilon_n) & \varpropto P(\varepsilon_n|A_n=i)P(A_n=i).
        \end{split}
    \end{equation}
In our experiments, we start with a uniform prior over sources and then prior information will be updated based on the observed EEG evidence as explained in \ref{Sec:OnlineControllerPerformance} as well.

\section{Data Collection and Preprocessing}
\label{cha:dataCollection}
\subsection{EEG Neurophysiological Data}
\label{cha:eegNeuralData}
Ten volunteers (5 male, 5 female), between the ages of 25 to 30 years, with no known history of hearing impairment or neurological problems participated in this study, which followed an IRB-approved protocol. EEG signals were recorded using a g.USBamp biosignal amplifier using active g.Butterfly electrodes with cap application from g.Tec (Graz, Austria) at 256 Hz. Sixteen EEG channels (P1, PZ, P2, CP1, CPZ, CP2, CZ, C3, C4, T7, T8, FC3, FC4, F3, F4 and FZ according to International 10/10 System) were selected to capture auditory related brain activities over the scalp. The selection was based on the topographical maps of classification performance observed and reported in our previous related work~\cite{haghighi2016audio}. Signals were filtered by built-in analog bandpass ([0.5, 60] Hz) and notch (60Hz) filters.

\subsection{Experimental Design} 
\label{cha:expDes}

Each participant completed one calibration and one online session of experiments.  Both sessions included diotic (both sounds playing on both ears simultaneously) auditory stimulation while the EEG was recorded from the participants. Participants passively listened to the auditory stimuli through earphones. 





\textbf{Calibration Session}\\
Total calibration session time was about 30 minutes. More specifically, a calibration session consisted of 60 trials of 20 seconds of diotic auditory stimuli with 4 seconds breaks between each trial. The diotic auditory stimuli are generated by one male and one female speaker. These speakers narrated a story (different story for different speakers chosen from audio books of literary novels) for 20 minutes. We consider every 20 seconds of this 20-minute-long diotic narration as a trial. During the calibration session, participants were asked to passively listen to 20-minute-long narration, and they were instructed to switch their attention from one speaker to another during different trials.  The instructions to switch attention from trial to trial are provided to the user on a computer screen using "f" and "m" symbols for female and male speakers, respectively.

\begin{figure}[h]
        \begin{minipage}[b]{1.0\linewidth}
            \centering
            \centerline{\includegraphics[width=13cm]{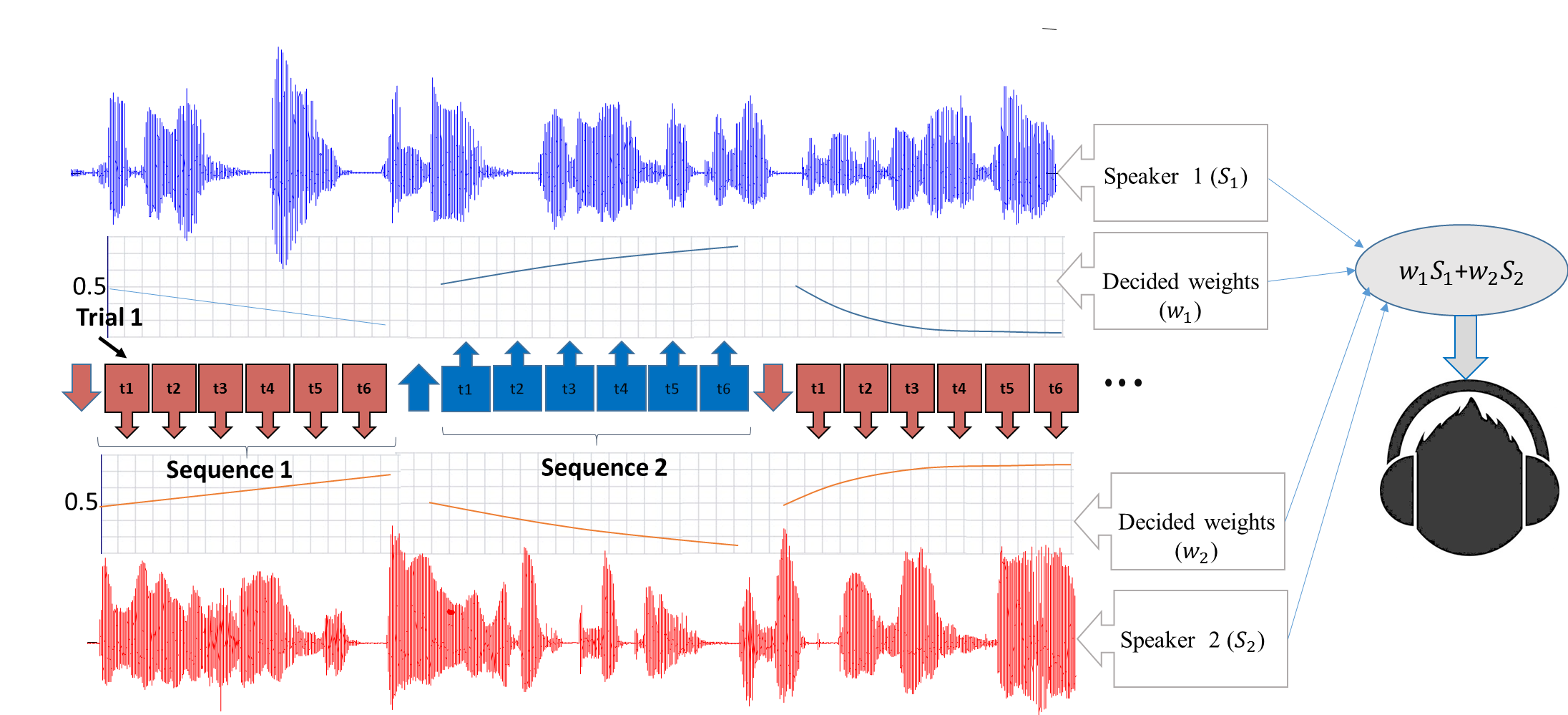}}
            \vspace{-15pt}
            \centerline{}\medskip
        \end{minipage}
        \caption{Online session experimental paradigm visualization. Two sounds are diotically playing in both ears. participants attend to the instructed sound source in each sequence. Each sequence starts with an equal weight trial and in its following trials weights get updated using attention inference and AGC modules.}
        \label{fig:expPara}
    \end{figure}

\textbf{Online Session}\\
The online session is summarized in Figure~\ref{fig:expPara}. In the online session, similar to the calibration session one male and one female speaker narrated stories (different story for different speakers) for 20 minutes. The same speakers from the calibration session narrated the stories for the online session, but the stories used in the online session were different than the calibration session. We consider the 20-minute-long narration as 10 two-minute long sequences, each sequence containing 6 twenty-second trials. Before each sequence, participants were asked to attend to one of the speakers through instructions displayed on the computer screen.  In each sequence, while the participants were listening to the narrated stories, weights that control the energy of each sound source were updated 6 times after every trial. The equal weight case is defined such that amplitude of each sound source was scaled to yield equal energy and each sequence started with an equal weight trial. 
There is a 0.5 second pause between 20-second-long trials within each sequence and the weights are updated within this 0.5 second period based on the attention evidence obtained from the EEG recorded from the participants and through the usage of automatic gain controller. Since, the participants were instructed to keep their attention on one of the speakers during each sequence, and during each sequence the weights are adjusted automatically in an online fashion to emphasize the attended sound source, we call this an online session.   
Silent portions of the story narration longer than 0.2 seconds were truncated to be 0.2 seconds, in order to reduce distraction of participants.

\subsection{Data Pre-processing}
\label{cha:preproc}

The acoustic envelope of speech stimulus signals were calculated using the Hilbert transform first and then both EEG brain activity measurements and speech envelopes were filtered by an FIR linear-phase bandpass filter ($[1.5,10]$Hz). Then, $t_x$ seconds of EEG and acoustic envelope signals following every stimulus and time locked to the stimulus onset were extracted.
Optimizing $t_x$  to get good performance with minimum time window is an important factor in the design of online auditory BCI systems. In this paper, we selected $t_x=20$, based on the results of our previous work which are reported in~\cite{haghighi2016audio}. The data length was selected based on the analysis we performed over the calibration data such that the length is chosen to optimize area under the receiver operating characteristics curve (AUC) of the intent inference engine with a constraint on the upper bound of the data length. More specifically, we analyzed the AUC as a function of the data length, and we chose the data length value when the changes in the AUC as the data length increased became more incremental for most of the participants.  

\section{Methods}
\label{cha:Methods}
\subsection{Feature Extraction}
\label{cha:FE}
Top down attention to an external sound source differentially modulates the neural activity to track the envelope of that sound source at different time lags ~\cite{ding2014cortical},~\cite{aiken2008human},~\cite{kong2014differential}. Therefore, as discriminative features, we calculate the cross correlation (CC) between the extracted EEG measurements and target and distractor acoustic envelopes at different time lags. $\boldsymbol{\tau_{n}}=[\tau_{1},~\cdots,~\tau_{l},~\cdots,~\tau_{L}]^\intercal$  is the vector of discreet time lag delays in sample between EEG and acoustic envelop of played sounds. In our analysis, we consider $\tau\in [t_1, t_2]\times{f_s}$, with $t_1$ and $t_2$ as sampling times chosen as described below. For each channel, we calculate cross correlations between EEG and the male and female speakers' acoustic envelopes for the time lag sample values defined in $\boldsymbol{\tau}$. Assuming that $\boldsymbol{\tau}$ is a  $L \times 1$ vector, we concatenate the cross correlation values from male and female speakers into a single vector and hence each feature vector is  $2L \times 1$ dimensional. $N$ is the number of EEG and sound source samples used for CC calculation. We have examined the effect of reducing $N$ on classification results in section \ref{cha:dataLenghPrf}.

Therefore, considering the defined notations, we calculate the correlation coefficient between EEG and sound sources at different time lag samples $\tau_{l}$, denoted by $\rho_{e^{ch},\hat{s}^{i}}{[\tau_l]}$:

\begin{equation}
   \rho_{e^{ch},\hat{s}^{i}}{[\tau_l]} = \frac{r_{e^{ch},\hat{s}^{i}}[\tau_l]}{\sigma_{e^{ch}},\sigma_{\hat{s}^{i}}}. \label{eqn:CC}
\end{equation}

In (\ref{eqn:CC}), $e^{ch}$ is EEG data recorded from channel $ch$, $\hat{s}^i$ is the envelope of $i^{th}$ estimated sound channel, $\tau_l$ is a time lag sample, and  ${r_{e^{ch},\hat{s}^{i}}[\tau_l]}=E[e_{n+\tau_l:N}^{ch},\hat{s}^{i}_{n:N-\tau_l}]$ is the sample average between $e^{ch}$ and $\hat{s}^i$. Therefore, $\rho_{e^{ch},\hat{s}^{i}}{[\tau_l]}$  is a scalar representing the correlation coefficient between EEG in channel $ch$ and $i^{th}$ sound channel at time lag sample $\tau_l$. 
 So,  $\mathbf{\rho_{e^{ch},\hat{s}^{i}}}=(\rho_{e^{ch},\hat{s}^{i}}{[\tau_1]},\rho_{e^{ch},\hat{s}^{i}}{[\tau_2]},...,\rho_{e^{ch},\hat{s}^{i}}{[\tau_L]})$ is $1 \times L$  dimensional vector for $L$ lags in $\tau$ range for channel $ch$ and $i^{th}$ sound channel. Feature vector will be formed by concatenation of correlation vectors for all $\hat{s}^{i}$'s. In our experiments which we have two sound sources this feature vector is specifically defined as $x^{ch}=[\mathbf{\rho_{e^{ch},\hat{s}^{1}}},\mathbf{\rho_{e^{ch},\hat{s}^{2}}}]^\intercal$. $x^{ch}$ is $2L \times 1$ vector for each channel and $\mathbf{x}=(x^{1},...,x^{ch},...,x^{16})^\intercal$ is a $2L \times 16$ dimensional matrix which contains features for each trial.

\subsection{Classification and Dimension Reduction}
\label{cha:Classification}
As explained in Section~\ref{cha:dataCollection}, the participants were asked to direct their auditory attention to a target speaker during data collection. The other speaker is the distractor. The labeled data collected in this manner is used in the analysis of discrimination between two speakers in a binary auditory attention classification problem. As explained in Section ~\ref{cha:FE}, for each trial we have $\mathbf{x}$ as the collection of $2L \times 1$ dimensional cross-correlation features for each channel. For analysis of data using all channels, we apply PCA first for dimensionality reduction to remove zero variance directions. Afterwards, feature vectors for each channel will be concatenated to form a single aggregated feature vector for further analysis.
Then, we use Regularized Discriminant Analysis (RDA)~\cite{friedman1989regularized} as the classifier in our analysis. RDA is a modification of Quadratic Discriminant Analysis (QDA). QDA assumes that data is generated by two Gaussian distributions with unknown mean and covariances and requires the estimation of these means and covariances of the target and nontarget classes before the calculation of the likelihood ratio. However, since, $L$, the length of $\boldsymbol{\tau}$, as defined in Section~\ref{cha:FE}, is usually large resulting in feature vectors with large dimensions even after the application of PCA, and the calibration sessions are short, the covariance estimates are rank deficient.


RDA eliminates the singularity of covariance matrices by introducing shrinkage and regularization steps. Assume each $\mathbf{x}_i \in \mathbb R^p$ is a $p \times 1$-dimensional feature vector for each trial and $y_i$ is its binary label showing if the feature belongs to speaker 1 or 2, that is $y_i \in \{1, 2\}$. Then the maximum likelihood estimates of the class conditional mean and the covariance matrices are computed as follows:

    \begin{equation}
        \begin{split}
            &\boldsymbol{\upmu}_k ={\frac{1}{N_k}}\sum_{i=1}^N
                {\mathbf{x}_i\delta(y_i,k)}, \\
            &\mathbf{\Sigma}_k ={\frac{1}{N_K}}\sum_{i=1}^N
                {(\mathbf{x}_i-\boldsymbol{\upmu}_k)
                (\mathbf{x}_i-\boldsymbol{\upmu}_k)^T\delta(y_i,k).} 
        \end{split}
    \end{equation}
where $\delta(\cdot,\cdot)$ is the Kronecker-$\delta$ function, $k$ represent a possible class label (here $k\in\{1,2\}$, and $N_k$ is the number of realizations in class $k$. Accordingly, the shrinkage and regularization of RDA is applied respectively as follows:
    \begin{equation}
        \begin{split}
            &\widehat{\mathbf{\Sigma}}_k(\lambda)=\frac{(1-\lambda)
                N_k\mathbf{\Sigma}_k+(\lambda)\sum_{k=0}^1{N_k\mathbf{\Sigma}_k}}
                {(1-\lambda) N_k+(\lambda)\sum_{k=0}^1{N_k}}, \\
            &\widehat{\mathbf{\Sigma}}_k(\lambda,\gamma)=
                (1-\gamma)\widehat{\mathbf{\Sigma}}_k(\lambda)+(\gamma) 
                {\frac{1}{p}}tr[\widehat{\mathbf{\Sigma}}_k(\lambda)]\mathbf{I}_{p}.
        \end{split}
    \end{equation}
Here, $\lambda,\gamma\in [0,1]$ are the shrinkage and regularization parameters, $tr[\cdot]$ is the trace operator and $\mathbf{I}_{p}$ is an identity matrix of size $p\times p$. In our system we optimize the values of $\lambda$ and $\gamma$ to obtain the maximum area under the receiver operating characteristics (ROC) curve (AUC) in a $5$-fold cross validation framework. Finally, the RDA score for a trial with the EEG evidence vector $\mathbf{x}_i$, which is defined as:
    \begin{equation}
        s_{\rm RDA}(\mathbf{x}_i)=\log\left(\frac{f_\mathcal{N}
        (\mathbf{x}_i;\boldsymbol{\upmu}_2,
        \widehat{\mathbf{\Sigma}}_2(\lambda,\gamma))}
        {f_\mathcal{N}(\mathbf{x}_i;\boldsymbol{\upmu}_1,
        \widehat{\mathbf{\Sigma}}_1(\lambda,\gamma))} \right),
    \end{equation}
where $f_\mathcal{N}(\mathbf{x};\boldsymbol{\upmu},\mathbf{\Sigma})$ is the Gaussian probability density function with mean $\boldsymbol{\upmu}$ and covariance $\mathbf{\Sigma}$. Here $s$ values are used to plot the ROC curves and to compute the AUC values. RDA can be considered as a nonlinear projection which maps EEG evidence to one dimensional score $\bm{\varepsilon}=s_{\scriptsize{\textrm{RDA}}}(\mathbf{x})$.

Finally, the conditional probability density function of $\bm{\varepsilon}$ given the class label, i.e. $p(\bm{\varepsilon}=\epsilon|A=i)$ needs to be estimated. We use kernel density estimation on the training data using a Gaussian kernel as
\begin{equation}
\hat{p}(\bm{\varepsilon}=\epsilon|A=i)=\frac{1}{N_i} \sum_{A(v)=i} K_{h_i}(\epsilon-\epsilon(v)),
\label{eq:conditionalpdfest}
\end{equation} 
where $\epsilon(v)$ is the discriminant score corresponding to a sample $v$ in the training data, that is to be calculated during cross validation, and $K_{h_k}(.)$ is the kernel function with bandwidth $h_k$. For a Gaussian kernel, the bandwidth $h_k$ is estimated using Silverman's rule of thumb ($\hat{h}_k=(\frac{4\hat{\sigma}^5}{3n})^{1/5}\cong 1.06\hat{\sigma}n^{-1/5}$) for each class $k$~\cite{silverman1986density}. This assumes the underlying density has the same average curvature as its variance-matching normal distribution~\cite{orhan2013offline}.

\section{Analysis and Results}
\label{cha:results}
As illustrated in our previous work,~\cite{haghighi2016audio}, features formed using the CC coefficient series  $\mathbf{\rho_{e^{ch},\hat{s}^{1}}},\mathbf{\rho_{e^{ch},\hat{s}^{2}}}$ as calculated in (\ref{eqn:CC}) show distinct patterns for attended vs unattended sound sources and these patterns are observed to be consistent across participants. For diotic presentation, the highest distinguishable absolute correlation between the sound sources and EEG is identified in the range of [0,400] ms. We accordingly extract features within this range of correlation delay, $\tau$. In this range, we observe a negative correlation for both target and distractor speakers followed by an early positive correlation for the target stimulus and delayed and suppressed version of that positive correlation for the distractor stimulus. These results are quantitatively summarized in Table \ref{tab:CCfeatures}, more specifically this table reports the average temporal latency and the magnitude of the peak in cross correlation responses across all participants. Statistical significance of the difference between peak temporal latency of target and distractor has been tested using Mann-Whitney U-test $(p=.00012)$.

    \begin{table}[h!]
        \centering
           \begin{adjustbox}{max width=13.0cm}
           \begin{tabular}{|c|c|c|c|}
            \hline
            Correlation Features    &  \bf{Positive Peak Magnitude Ratio}& \multicolumn{2}{c|}{\bf{Time Lag of Peak (ms)}}  \\ \hline
                 Stimulus          &Target / Distractor  &    Target &	Distractor \\ \hline
                 Average for all Participants (mean $\pm$ sd)          & $2.08\pm 1.1$ &	$159.34\pm 11$ &	$225.78\pm 42.9$ \\
                \hline
            \end{tabular}
            \end{adjustbox}
        \caption{Average of time latency and magnitude of peak in cross correlation responses across all participants.}
        \label{tab:CCfeatures}
    \end{table}
    In the rest of the analysis, we consider the correlation delay $\tau$ to be in the range of [0,400]ms to form the feature vectors.     

\subsection{Offline Data Analysis}
\subsubsection{\textbf{Single channel classification analysis}}

\begin{figure}[h]
    \begin{minipage}[b]{1.0\linewidth}
    \centering
    \centerline{\includegraphics[width=11cm]{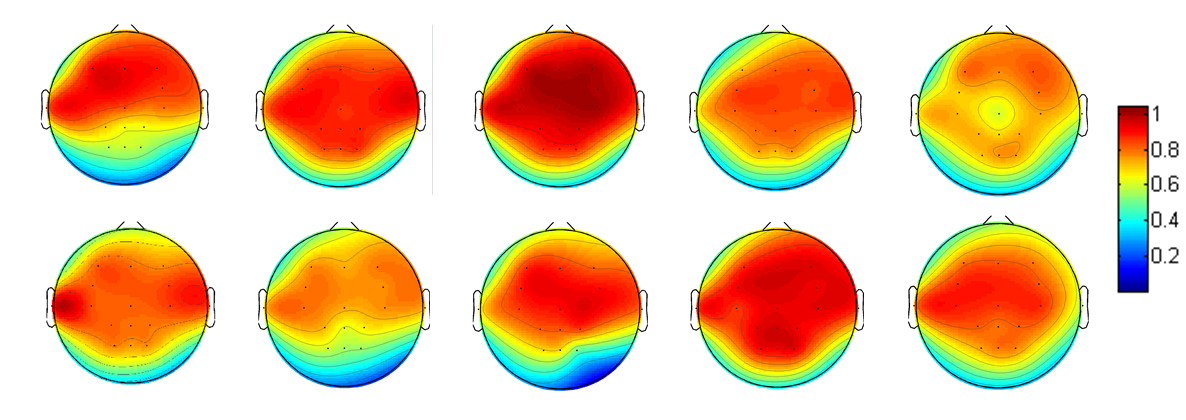}}
    \vspace{-15pt}
    \centerline{}\medskip
    \end{minipage}      
    \caption{Topographic map of classification performance over the scalp for classifying attended versus unattended speakers for all participants.}
    \label{fig:topomaps}
\end{figure}

Using the selected window of [0,400] ms as the most informative window for classification of target versus distractor responses, we first form the vector $x^{ch}$ as shown in \ref{cha:FE}, we then use these features for each EEG channel independently to localize the selective attention responses using the classification scheme described in Section~\ref{cha:Classification}. As the results of our previous work suggested~\cite{haghighi2016audio}, we relocated electrodes to be more centered around the frontal cortex, see Section~\ref{cha:eegNeuralData}. Figure~\ref{fig:topomaps} shows the topographical map of classification performance in terms of area under the receiver operating characteristics curve (AUC) over the scalp, for all participants. Moreover, for each participant best channel AUC values are reported in Table~\ref{tab:maxPerfChannel}. Figure~\ref{fig:topomaps} and Table~\ref{tab:maxPerfChannel} show that the classification accuracy varies across participants, but for each participant channels located in central and frontal cortices have higher classification accuracy. 

\begin{table}[h!]
        \centering
    \begin{adjustbox}{max width=12.0cm}
        \begin{tabular}{|l||*{10}{c|}}\hline
            \makebox{Participant}
            &\makebox[3em]{1}&\makebox[3em]{2}&\makebox[3em]{3}&\makebox[3em]{4}&\makebox[3em]{5}
            &\makebox[3em]{6}&\makebox[3em]{7}&\makebox[3em]{8}&\makebox[3em]{9}&\makebox[3em]{10}\\\hline\hline
            Best AUC  &0.92&0.92&1&0.84&0.83&0.92&0.80&0.91&0.96&0.89\\\hline
            Best channel &Fz&C3&	C4&	Fc3&F4&	T7&	C3&	C4&	CPz&C3\\\hline
        \end{tabular}
    \end{adjustbox}%
    \caption{Channel with maximum performance and its corresponding AUC performance.}
\label{tab:maxPerfChannel}
\end{table}

\subsubsection{\textbf{Classification performance versus trial length analysis}}
\label{cha:dataLenghPrf}
\begin{figure}[h]
        \begin{minipage}[b]{1.\linewidth}
        \centering
        \centerline{\includegraphics[width=14cm]{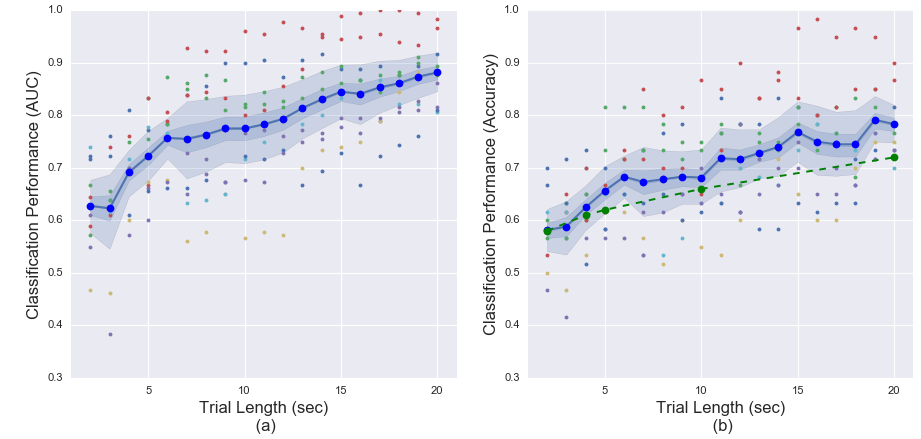}}
        \vspace{-15pt}
        \centerline{}\medskip
        \end{minipage}
        \caption{Performance versus trial length curves considering (a) AUC and (b) Accuracy as performance metrics. Different colored dots are used to represent the performance of  different participants. The blue curve is the average of performance values over all 10 participants. The green line presents the performance results of a previous approach.}
        \label{fig:perfVsDataLength}
    \end{figure}

In this section we analyze the effect of trial length on classification performance. Specifically, using the calibration data, we consider different lengths (from 2 seconds to 20 seconds) of EEG and estimated sound sources to calculate the cross correlations and extract features accordingly to train our classifier to distinguish the attended sound source from the unattended one. Figure \ref{fig:perfVsDataLength} shows the classification performance using all 16 channels. In this figure, different colors represent the performances of different participants. The blue curve is the average of performance over all 10 participants using different data lengths for classification. Dark and light shaded areas around the average line shows the 50 and 95 percent confidence interval calculated according to the bootstrap method, respectively. Figure~\ref{fig:perfVsDataLength}(a) shows AUC performance while Figure~\ref{fig:perfVsDataLength}(b) shows probability of correct decision (i.e., accuracy). Moreover, Figure~\ref{fig:perfVsDataLength}(b) also compares our results with a related previous work that is presented in \cite{horton2014envelope}. The performance reported for 128 channels in that previous work is illustrated as a green line in this figure. In this figure, we observe that using much smaller number of channels, our method outperforms the previous approach.

\subsection{Online Controller Performance}
\label{Sec:OnlineControllerPerformance}
Recall that as explained in Section~\ref{cha:expDes}, the online experiment includes listening to 10 two-minute sequences. During each sequence the participants were requested to focus their auditory attention to one of the speakers. Each sequence contains multiple trials and within each sequence we perform adaptive sound source weight estimation and update after every trial (20 seconds). More specifically, we calculate the EEG evidence as explained in Section~\ref{cha:FE}. Using conditional probability density functions as described in section \ref{cha:Classification}, we obtain the posterior estimate of the probability for each class being the intended source, which is proportional to class conditional likelihoods times prior knowledge on probability of attention. Then source weights for each source are adjusted as being proportional to the posterior probability of that class given EEG evidence. 




\begin{align} \label{eq8}
    w_{k,n+1}^{(i)}\enskip &= p(A_{k}=i|\varepsilon_{k,n}, \ldots, \varepsilon_{k,1})\\ 
    &= \enskip \frac{p(A_{k}=i|\varepsilon_{k,n}) p(A_{k}=i|\varepsilon_{k,n-1}, \ldots, \varepsilon_{k,1})}{\displaystyle\sum_{j=1}^2p(A_{k}=j|\varepsilon_{k,n})p(A_{k}=j|\varepsilon_{k,n-1}, \ldots, \varepsilon_{k,1})} \quad,i=1,2 \\
    &=\enskip \frac{p(A_{k}=i|\varepsilon_{k,n})\cdot w_{k, n}^{(i)}}{\displaystyle\sum_{j=1}^2p(A_{k}=j|\varepsilon_{k,n})\cdot w_{k, n}^{(j)}} \quad,i=1,2.
\end{align}

\begin{figure}[h!]
        \begin{minipage}[b]{1.0\linewidth}
        \centering
        \centerline{\includegraphics[width=15cm]{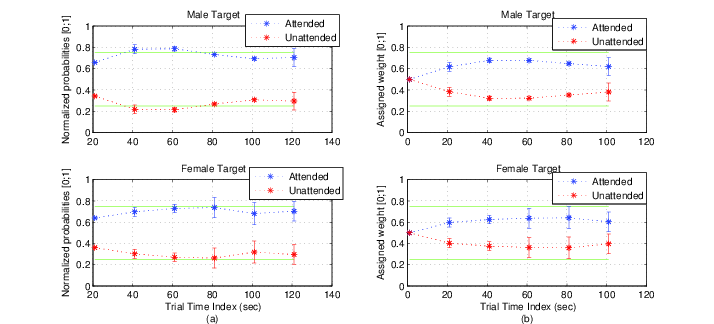}}
        \vspace{-15pt}
        \centerline{}\medskip
        \end{minipage}
        \vspace{-1.2cm}
        \caption{(a) Normalized posterior probabilities which are computed after every trial, (b) Weights assigned to each trial at the beginning of that trial which are calculated based on normalized probabilities of their preceding trial.Values in both figures are averaged over all trials and participants for female and male target separately.}
        \label{fig:onlineRes}
    \end{figure}
\begin{figure}[h!]
        \begin{minipage}[b]{1.0\linewidth}
        \centering
        \centerline{\includegraphics[width=15cm]{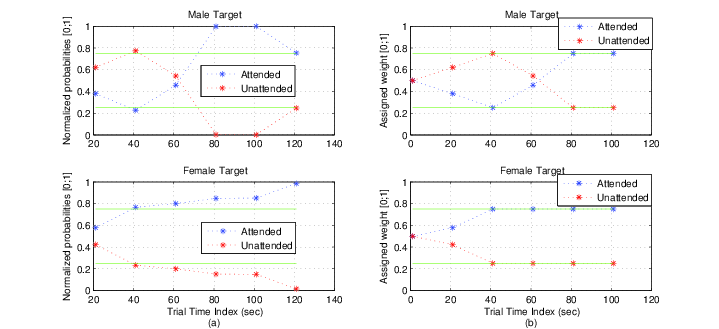}}
        \vspace{-15pt}
        \centerline{}\medskip
        \end{minipage}
        \vspace{-1.2cm}
        \caption{Two examples for a normal versus mistake recovery case. (First row) is showing an example of a sequence with mistake recovery by participant. Left column is showing the calculated normalized probabilities at the end of each trial and right column is showing the corresponding assigned weights to normalized probabilities at the beginning of each trial. (Second row) in an example of a normal sequence. Limits imposed on the assigned weights are shown with green constant lines in all figures.}
        \label{fig:mistakeCorrectionRes}
    \end{figure}

In the equation above, $k$ is the sequence index and $n$ is the trial index. Each sequence contains 6 trials and during each sequence we assume that the user is focusing on the same sounds source. This equation assumes that the attention remains on the same source during the updates in each sequence. Also in this weight update equation above we initialize $p(A_{k}=i|\varepsilon_{k,0})=0.5$.  We trained the system using a calibration session and tested the learned model in an online session. Users attempted to amplify the designated target speech with their auditory attention using this brain interface in 10, two-minute-long sequences. Figure \ref{fig:onlineRes} shows the average of decided weights (at every 20 seconds over 5 trials) for attended and unattended speech sources over the course of two minutes, for male and female narrators. Figure \ref{fig:onlineRes} (a) is showing the average of the estimated probabilities for each class at the end of each trial using its preceding 20 seconds of data, as stated in equation \ref{eq8}. Figure \ref{fig:onlineRes} (b) shows the average of employed weights instead of normalized probabilities. The difference between Figures \ref{fig:onlineRes} (a) and (b) is due to the limits imposed on weights ([0.25 to 0.75] which are shown with green constants). 
These limitations were imposed to ensure the audibility of both sources, to enable mistake correction in the event of algorithm/human errors, and to allow shifting attention if desired. Figure~\ref{fig:mistakeCorrectionRes} illustrates two example sequences:  one for a normal case in which there is no algorithm/human error (second row of the figure), and the first row of the figure demonstrates a case in which a participant is able to recover from a potential error in detecting the attended sound source. In this second case, the weight of the attended sound source was lower than the unattended one; however, since the system imposes a lower bound on the weights, the participant was able to recover and the weight of the attended sound source increased accordingly before the sequence ended.   
 
\subsection{Online Vs Offline data analysis}
Since changes in energy and amplitude of competing sound sources will potentially change the statistics of the EEG measurements, analyzing how robust feature vectors are to these changes can help us understand the impacts of the weights of the sound sources on the attention and EEG models. Table \ref{tab:onlineOfflineAUCs} shows the generalization of the classifier trained on the calibration data and tested on the data collected during the online sessions when the EEG from all the channels were used. Specifically, the first and the second rows of the table present the AUC results when 5-fold cross validation is performed on the calibration and online session data, respectively. The results in the third row are obtained when the classifier is trained using the calibration data and tested on the online data. Therefore, the third row demonstrates the generalization of the trained auditory attention classifier from the calibration session to the online testing. Note here that the third row demonstrates the performance of the auditory attention classifier when it was used in online session where the sound source weights were adaptively updated. From this table, we observe that there is a decrease in the performance when the classifier is used in the online session compared to the calibration session. Even though the classification accuracy is acceptable when the classifier trained on the calibration data and tested on the online data as illustrated in row 3 of the table, a calibration session with varying weights on the sounds sources could potentially improve the classification accuracy further. This will be the focus of our future work.


\begin{table}[h!]
        \centering
    \begin{adjustbox}{max width=12.0cm}
        \begin{tabular}{|l||*{10}{c|}}\hline
            \diagbox{AUC}{Participant}
            &\makebox[3em]{1}&\makebox[3em]{2}&\makebox[3em]{3}&\makebox[3em]{4}&\makebox[3em]{5}
            &\makebox[3em]{6}&\makebox[3em]{7}&\makebox[3em]{8}&\makebox[3em]{9}&\makebox[3em]{10}\\\hline\hline
            Calibration Data (offline)  &0.91&0.89&1&0.81&0.88&0.82&0.77&0.89&0.94&0.83\\\hline
            Online Session Data &0.83&0.74&0.98&0.63&0.82&0.8&0.74&0.92&0.83&0.69\\\hline
            Calibration Model on Online Data &0.86&0.77&0.95&0.73&0.77&0.83&0.8&0.82&0.86&0.70\\\hline
        \end{tabular}
    \end{adjustbox}
        \caption{AUCs for offline and online data independently and applying the learned model from offline data on online data.}
\label{tab:onlineOfflineAUCs}
\end{table}

\section{Conclusion, Limitations and future work}
This work is a novel attempt to investigate the feasibility of a close loop online sound source modulation system using a non-invasive EEG-based brain interface. In a scenario to detect the attended sound source in the presence of two speakers, we presented two main contributions in this paper. First, we showed high offline attended sound source classification accuracy with single channel EEG when the EEG duration was 20 seconds. Second, the novel brain interface presented in this manuscript utilizes an automatic gain control to adjust the amplitudes of attended and unattended sound sources with the goal of increasing signal-noise-ratio and improving listening and hearing comfort. Through an experimental study, we showed that the designed BCI together with the automatic gain control has the potential to improve the information rate by reducing the trial lengths and increasing the classification accuracies for shorter trial lengths compared to the performance results reported in the existing related works. Even though promising results were obtained with this proof of concept study, there are many opportunities to improve the performance of the system. For example, various different techniques could be investigated to optimize the automatic gain control scheme or the classification method with the purpose of enabling fast and accurate decision making in an online setting. This improvement is essential for the presented BCI to be a practical reality and potentially be a part of the future generations of hearing aids. 


\section*{Acknowledgment}
This work is supported by NSF (CNS-1136027, IIS-1149570, CNS-1544895), NIDLRR (90RE5017-02-01), and NIH (R01DC009834).



\section*{References}
\bibliography{refs}

\end{document}